\documentclass[12pt, aps, prd, reprint, sort&compress, superscriptaddress, showpacs, nofootinbib, preprintnumbers]{revtex4-1}
\pdfoutput=1
\usepackage{graphicx}
\usepackage{dcolumn}
\usepackage{bm}
\usepackage{amsmath}
\usepackage{amssymb}
\usepackage{dsfont}
\usepackage{amsfonts}
\usepackage[UKenglish]{babel}
\usepackage{hyperref}
\usepackage{nicefrac}
\usepackage{verbatim}
\usepackage{bbm}
\usepackage{color}

\begin{document}

\title{Exploring a new $SU(4)$ symmetry of meson interpolators}
\date{\today}
\author{L.~Ya.~Glozman}
\email{leonid.glozman@uni-graz.at}
\author{M.~Pak}
\email{markus.pak@uni-graz.at}
\affiliation{Institut f\"ur Physik, FB Theoretische Physik, Universit\"at Graz, Universit\"atsplatz 5,
8010 Graz, Austria}

\begin{abstract}
In recent lattice calculations it has been discovered 
that  mesons upon truncation of the quasi-zero modes of the Dirac
operator obey a symmetry larger than  the $SU(2)_L \times SU(2)_R\times U(1)_A$ symmetry of the QCD Lagrangian. 
This symmetry has been suggested to be
$SU(4) \supset SU(2)_L \times SU(2)_R\times U(1)_A$ that
mixes not only the u- and d-quarks of a given chirality, but also
the left- and right-handed components. Here it is demonstrated that 
bilinear $\overline{q}q$ interpolating fields of a given spin $J \geq 1$ transform into
each other according to irreducible representations of  $SU(4)$
or, in general, $SU(2N_F)$. 
This fact together with the coincidence of the correlation functions establishes
$SU(4)$ as a symmetry of the $J \geq 1$ mesons upon quasi-zero mode
reduction. It is shown that this symmetry  is a symmetry of the confining
instantaneous charge-charge interaction in QCD.
Different subgroups of $SU(4)$ as well as the $SU(4)$ algebra are explored.  
\end{abstract}


\maketitle

\section{Introduction}
In recent $N_{\textsc{F}}=2$ dynamical lattice simulations with the manifestly
chiral-invariant Overlap Dirac operator, a new symmetry of mesons of 
given spin has been discovered upon truncation of the quasi-zero modes
of the Dirac operator, Refs.~\cite{DGL1, DGL2} (A hint
for this symmetry had been seen in a previous study, Ref.~\cite{GLS}).
Namely, the $J=1$ mesons $\rho,\rho',\omega,\omega',a_1,b_1,h_1,f_1$ get
degenerate after removal of the lowest-lying Dirac eigenmodes\footnote{It is not yet entirely clear from the lattice results whether the $f_1$ state
is degenerate with other $J=1$ mesons. While the quality of the effective mass
plateau is excellent for the $\rho,\rho',\omega,\omega',a_1,b_1,h_1$ mesons
it is not so for the $f_1$ state.}. A similar
degeneracy is seen also in $J=2$ mesons, Ref.~\cite{DGLP}. This symmetry has been
suggested to be $SU(4) \supset SU(2)_L \times SU(2)_R\times U(1)_A$ that
mixes components of the fundamental vector $(u_L,u_R,d_L,d_R)$, Ref.~\cite{G1}.
It is higher than the broken $SU(2)_L \times SU(2)_R\times U(1)_A$
symmetry of the QCD Lagrangian and  should be considered as
an emergent symmetry  in $J \geq 1$ mesons that reflects the
QCD dynamics once the quasi-zero
modes of the Dirac operator have been removed.  It has been proposed that this symmetry
might be a symmetry of the dynamical QCD string because there is no color-magnetic interaction (field) in the system, Ref.~\cite{G1}.

In the present paper we extend findings of the Letter \cite{G1} and show that 
the  composite $J \geq 1$ $\overline{q} q$ bilinear operators (interpolating fields) with non-exotic quantum numbers transform
according to irreducible dim$=15$ and dim$=1$ representations of the $SU(4) \supset SU(2)_L \times SU(2)_R\times U(1)_A \times C_i$ group. 
This result holds irrespective of the observations made in Refs.~\cite{DGL1,DGL2} as well as possible physics interpretations in Ref.~\cite{G1}. 
The correlation functions obtained with these operators get indistinguishable after truncation, Ref.~\cite{DGL2}. This fact establishes
consequently the proposed $SU(4)$ symmetry as the symmetry of the $J \geq 1$ spectra upon the quasi-zero mode reduction. We also study different subgroups
of $SU(4)$, the corresponding algebras as well as transformation properties of the interpolators with respect to these subgroups.

The outline of the article is as follows: In Chapter II we review the classification of the spin-1 $\overline{q} q$-bilinears with respect
to $SU(2)_L \times SU(2)_R$ and $U(1)_A$ transformations, Ref.~\cite{CJ}.
In Chapter III we demonstrate that all these interpolators are connected with each other through the $SU(4)$ transformations
that include not only the chiral rotations but also a mixing between the
left- and right-handed components, specify interpolators that
transform according to different subgroups of $SU(4)$ and construct the
respective algebras. A generalization to $SU(2N_F)$ and to general spin is also discussed. In the last Chapter we show that the observed $SU(4)$ symmetry implies the
absence of the color-magnetic field in the confined system after the low-mode elimination  and can be considered as a manifestation of the dynamical string
in QCD.

\section{Chiral classification of the $J=1$ bilinear operators.}
\label{Meson-class}
\begin{figure}
\centering
\includegraphics[angle=0,width=.95\linewidth]{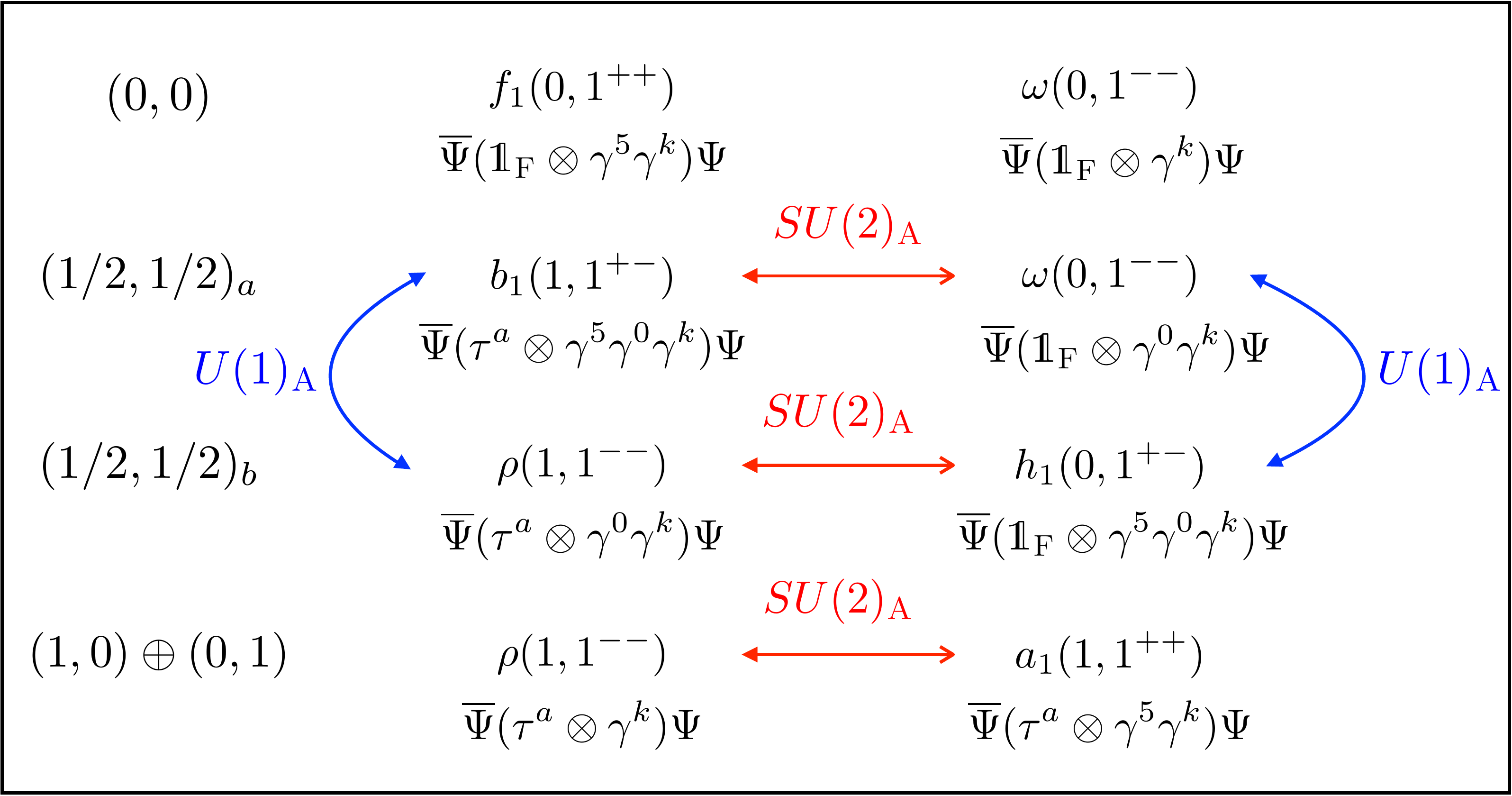}
\caption[Chiral-Parity group 1]{\sl On the left column the irreducible representations of the parity-chiral group are given. Each 
meson is denoted as $(I, J^{PC})$, with $I$ isospin, $J$ total spin, $P$ parity and $C$ charge conjugation.
Below each state a current from which it can be generated, is given. The $SU(2)_A$ and $U(1)_A$ connections are denoted by red and blue lines, respectively.}
 \label{Table1}
\end{figure}

We work in Minkowski space with the chiral representation of the $\gamma$-matrices. In flavor space 
we use the Pauli matrices $\boldsymbol{\tau}$. The 
basic definitions are collected in Appendix \ref{App-def}. 
With the notation
\begin{align}
\label{Psi}
 \Psi= \begin{pmatrix}
  u \\
  d  
 \end{pmatrix}  
\end{align}
we make explicit the two flavors in the quark field.
The left- and right-handed quark fields for one flavor are defined 
via the projection operator $P_{\pm} = 1/2 (\mathds{1}\pm \gamma^5)$, 
which can be generalized for two quark flavors by defining the 
projectors as $\Gamma_{\pm} = (\mathds{1}_F \otimes P_{\pm})$:
\begin{align}
 \Psi_{\textsc{L}} = \Gamma_{-} \Psi \; ,  \Psi_{\textsc{R}} = \Gamma_{+} \Psi \; . 
\end{align}
All $\overline{q} q$-mesons and respective operators
with non-exotic quantum numbers can be arranged into irreducible 
representations of the parity-chiral group $SU(2)_L \times SU(2)_R \times C_i$, Ref.~\cite{G2}. 
We use the notation $(I_{\textsc{L}}, I_{\textsc{R}})$, with left-handed ($I_{\textsc{L}}$)
and right-handed ($I_{\textsc{R}}$) isospin for each irreducible representation of $SU(2)_L \times SU(2)_R$. The classification of spin-1 meson
operators is presented in Fig.~\ref{Table1}. 
Below each meson a corresponding 
interpolator $J_{(I,J^{PC})}^r$ is given with $r$ being the index of an irreducible representation of the parity-chiral group. 
 
As an example we now compare the combination of left- and right-handed quarks within the interpolators of the two isovectors $1^{--}$.
We start with the interpolators $J^{(1,0) \oplus (0,1)}_{(1,1^{--})} = \overline{\Psi} (\tau^a \otimes \gamma^k) \Psi$ 
and write it in terms of left- and right-handed quarks:
\begin{align}
\label{curr1}
 J_{(1,1^{--})}^{(1,0) \oplus (0,1)} = \overline{\Psi}_{\textsc{L}} (\tau^a \otimes \gamma^k) \Psi_{\textsc{L}} +  \overline{\Psi}_{\textsc{R}} (\tau^a \otimes \gamma^k) \Psi_{\textsc{R}} \; .
\end{align}
It has the chiral content $\overline{L} L + \overline{R} R$. 
The interpolator $J^{(1/2,1/2)_b}_{(1,1^{--})} = \overline{\Psi} (\tau^a \otimes \gamma^0 \gamma^k) \Psi$ can be split up as
\begin{align}
\label{curr2}
 J^{(1/2,1/2)_b}_{(1,1^{--})} = \overline{\Psi}_{\textsc{L}} (\tau^a \otimes \gamma^0 \gamma^k) \Psi_{\textsc{R}} +  
 \overline{\Psi}_{\textsc{R}} (\tau^a \otimes \gamma^0 \gamma^k) \Psi_{\textsc{L}} \; ,
\end{align}
and has the chiral content $\overline{L} R + \overline{R} L$. 

The axial part of the $SU(2)_L \times SU(2)_R$ transformations is defined
by
\begin{align}
  \Psi \rightarrow  \Psi^{'} = e^{i \frac{\boldsymbol{\varepsilon} \cdot \boldsymbol{\tau}}{2} \otimes \gamma^5} \Psi \equiv U \, \Psi \; .
\end{align}
\noindent
These axial transformations do not form a closed group. However, we use $SU(2)_A$
as a shorthand notation for these transformations in the text below and 
in  Fig.~\ref{Table1}.

The matrix $U$ has the property $U^{\dagger} (\mathds{1}_{\textsc{F}} \otimes \gamma^0) = (\mathds{1}_{\textsc{F}} \otimes \gamma^0) U$, from which
$\overline{\Psi}' = \overline{\Psi} U$ follows. It can be expressed in closed form as 
\begin{align}
\label{U-closed}
U = (\mathds{1}_{\textsc{F}} \otimes \mathds{1}_{\textsc{D}}) \cos\left[\frac{|\boldsymbol{\varepsilon}|}{2}\right]
 + i ( \hat{\boldsymbol{\varepsilon}} \cdot \boldsymbol{\tau} \otimes \gamma^5) \sin\left[\frac{|\boldsymbol{\varepsilon}|}{2}\right] \; , 
\end{align}
with $\hat{\boldsymbol{\varepsilon}} = \boldsymbol{\varepsilon}/|\boldsymbol{\varepsilon}|$.
We now apply the  $ SU(2)_A$ transformation $U$ on the individual interpolators
of Fig.~\ref{Table1}. For instance, the interpolator $J_{(0,1^{--})}^{(1/2,1/2)_a}$ transforms as 
\begin{align}
\overline{\Psi}' (\mathds{1}_{\textsc{F}} \otimes \gamma^0 \gamma^k) \Psi' = 
&\overline{\Psi} (\mathds{1}_{\textsc{F}} \otimes \gamma^0 \gamma^k) \Psi \cdot \mathcal{E}  + \nonumber \\
&\overline{\Psi} (\tau^a \otimes \gamma^5 \gamma^0 \gamma^k) \Psi \cdot \mathcal{F}^a \; , 
\end{align}
with $\mathcal{E} = \cos |\boldsymbol{\varepsilon}|$ and
$\mathcal{F}^a = i \hat{\boldsymbol{\varepsilon}}^a \sin|\boldsymbol{\varepsilon}|$
being functions of the rotation vector $\boldsymbol{\varepsilon}$ only.
We find that the following
pairs become connected via the $SU(2)_A$ (see Fig.~\ref{Table1}):
\begin{align}
\label{rel-a}
J_{(1,1^{+-})}^{(1/2,1/2)_a} &\longleftrightarrow J_{(0,1^{--})}^{(1/2,1/2)_a} \; , \\
J_{(1,1^{--})}^{(1/2,1/2)_b} &\longleftrightarrow J_{(0,1^{+-})}^{(1/2,1/2)_b} \; ,  \\
J_{(1,1^{--})}^{(1,0) \oplus (0,1)} &\longleftrightarrow J_{(1,1^{++})}^{(1,0) \oplus (0,1)} \; .
\end{align}
\noindent
Similarly, the $U(1)_A$ transformation
\begin{align}
  \Psi \rightarrow  \Psi^{'} = e^{i \alpha (\mathds{1}_{\textsc{F}} \otimes \gamma^5)} \Psi \; , 
\end{align}
connects interpolators from the $(1/2,1/2)_a$ and $(1/2,1/2)_b$ 
representations which have the same isospin but opposite spatial parity. These
four interpolators form an irreducible representation of $SU(2)_L \times SU(2)_R \times U(1)_A$. 
The interpolators from the  $(1,0) \oplus (0,1)$
representation are self-dual with respect to the $U(1)_A$ transformations.
The singlet interpolators from the $(0,0)$ representations are invariant
with respect to both $U(1)_A$ and $SU(2)_A$ transformations.

\section{Extending $SU(2)_L \times SU(2)_R \times U(1)_A \times C_i$ to $SU(4)$}

\subsection{Left-right mixing}
Our task  is to find transformations that mix
different representations of the parity-chiral group.
The representations  $(1/2,1/2)$
have the quark content $\overline{L} R \pm \overline{R} L$ and the representations $(0,0), (1,0)\oplus(0,1)$ have the quark content 
$\overline{L} L \pm \overline{R} R$. Consequently, in order to connect these
representations one needs to find
a symmetry transformation, which mixes left- and right-handed quarks, Ref.~\cite{G1}.

Consider the fundamental doublets $\textsc{U} =\begin{pmatrix} u_L \\ u_R \end{pmatrix} $ 
and $\textsc{D} =\begin{pmatrix} d_L \\ d_R \end{pmatrix} $ constructed from Weyl
spinors. We can consider $SU(2)_{\textsc{u}}$ and $SU(2)_{\textsc{d}}$ 
rotations of these doublets in an imaginary three-dimensional space that mix the $u_{\textsc{L}}$ and
$u_{\textsc{R}}$ as well as the $d_{\textsc{L}}$ and $d_{\textsc{R}}$ spinors. It is similar to the well familiar
concept of the isospin space: The electric charges of particles are conserved
quantities, but rotations in the isospin space mix particles with different
electric charges. In our case the chirality of a massless quark is a conserved
quantity but the $SU(2)_{\textsc{u}}$ and $SU(2)_{\textsc{d}}$ rotations mix quarks with different
chiralities:
\begin{align}
\label{chiralspin}
  \textsc{U} \rightarrow  \textsc{U}{'} = e^{i \frac {\boldsymbol{\varepsilon} \cdot \boldsymbol{\sigma}}{2}} \textsc{U}\; ,~~~~~~
 \textsc{D} \rightarrow  \textsc{D}{'} = e^{i \frac {\boldsymbol{\varepsilon} \cdot \boldsymbol{\sigma}}{2}} \textsc{D} \; ,
\end{align}
where $\boldsymbol{\sigma}$ are the standard Pauli matrices which obey
the $\mathfrak{su}(2)$ algebra:
\begin{align}
 [\sigma^i,\sigma^j] = 2 i \epsilon^{i j k} \, \sigma^k \; .
\end{align}
We refer to this imaginary three-dimensional space as the \textit{chiralspin} space.

Instead of the Weyl spinors we can consider the left- and right-handed
Dirac bispinors. Then, the required $\mathfrak{su}(2)$ algebra can
be constructed with the $4 \times 4$ matrices
\begin{align}
\label{sigma}
\boldsymbol{\Sigma} = \{ \gamma^0, i \gamma^5 \gamma^0, -\gamma^5 \} \; ,  
\end{align}
with the commutation relation
\begin{align}
 [\Sigma^i,\Sigma^j] = 2 i \epsilon^{i j k} \, \Sigma^k \; .
\end{align}
These rotations act in Dirac space only and are diagonal in flavor space:
\begin{align}
\label{V-def}
  \Psi \rightarrow  \Psi^{'} &= e^{i (\mathds{1}_{\textsc{F}} \otimes \frac {\boldsymbol{\varepsilon} \cdot \boldsymbol{\Sigma}}{2})} \Psi \equiv V \, \Psi \; .
\end{align}
\noindent
We denote this symmetry group as $SU(2)_{\textsc{cs}}$. We note that in the compact notation
of Eq.~(\ref{V-def}) two $SU(2)_{\textsc{u}}$ and $SU(2)_{\textsc{d}}$ symmetries for the individual quark flavors $u$ and $d$ are hidden.

In analogy to Eq.~(\ref{U-closed}) we express $V$ as 
\begin{align}
\label{V-closed}
 V = (\mathds{1}_{\textsc{F}} \otimes \mathds{1}_{\textsc{D}}) \cos\left[\frac{|\boldsymbol{\varepsilon}|}{2}\right]
 + i (\mathds{1}_F \otimes \hat{\boldsymbol{\varepsilon}} \cdot \boldsymbol{\Sigma}) \sin\left[\frac{|\boldsymbol{\varepsilon}|}{2}\right] \; . 
\end{align}
Now we apply these chiralspin rotations on the interpolators in Fig.~\ref{Table1}
and find the following triplets\footnote{Consequently, the chiralspin 1
should be ascribed to these fields.} of interpolators
that are connected to each other:\footnote{When applying $V$ on $\overline{\Psi}$ we have to be careful, since $V$ and $(\mathds{1}_{\textsc{F}} \otimes \gamma_0)$ do not commute.
We write $\overline{\Psi}' = \overline{\Psi} (\mathds{1}_{\textsc{F}} \otimes \gamma^0) V^{\dagger} (\mathds{1}_{\textsc{F}} \otimes \gamma^0)$.}
\begin{align}
\label{V-mes1}
 J_{(0,1^{--})}^{(0,0)} &\leftrightarrow J_{(0,1^{--})}^{(1/2,1/2)_a}  \leftrightarrow J_{(0,1^{+-})}^{(1/2,1/2)_b} \; , \\
\label{V-mes2}
 J_{(1,1^{--})}^{(1,0) \oplus (0,1)} &\leftrightarrow J_{(1,1^{--})}^{(1/2,1/2)_b} \leftrightarrow J_{(1,1^{+-})}^{(1/2,1/2)_a} \; .
\end{align}
This means, that transforming any of the interpolators in Eq.~(\ref{V-mes1}) with respect to $V$,
the result can always be decomposed as
\begin{align}
 \overline{\Psi} (\mathds{1}_{\textsc{F}} \otimes \gamma^k) \Psi \cdot &\mathcal{E}^{(i)} + \overline{\Psi}(\mathds{1}_{\textsc{F}} \otimes \gamma^5 \gamma^0 \gamma^k) \Psi \cdot \mathcal{F}^{(i)}  \nonumber \\
&+ \overline{\Psi} (\mathds{1}_{\textsc{F}} \otimes  \gamma^0 \gamma^k) \Psi \cdot \mathcal{G}^{(i)} \; , 
\end{align}
with $i=1,2,3$ labeling the interpolators, and $\mathcal{E}^{(i)}, \mathcal{F}^{(i)}, \mathcal{G}^{(i)}$ being functions of the rotation vector
$\boldsymbol{\varepsilon}$ only. Performing a transformation of the interpolating
currents in Eq.~(\ref{V-mes2}) leads to the same decomposition with $\tau^a$ instead of $\mathds{1}_{\textsc{F}}$ in flavor space. 
It is clear, why we get two triplets of states: in Eq.~(\ref{V-def}) two $SU(2)$ symmetries, namely for up and down quarks,
appear
\footnote{The symmetry $SU(2)_{CS}$ connects the interpolators with off-diagonal $\gamma$-structure to
interpolators with diagonal $\gamma$-structure (in the chiral representation of the $\gamma$-matrices). This is how left- and right-handed
quarks are mixed.}.
The interpolators 
\begin{align}
\label{invar1}
 J^{(0,0)}_{(0,1^{++})} &= \overline{\Psi} (\mathds{1}_{\textsc{F}} \otimes \gamma^5 \gamma^k) \Psi \; , \\
  J^{(1,0) \oplus (0,1)}_{(1,1^{++})} &= \overline{\Psi} (\tau^a \otimes \gamma^5 \gamma^k) \Psi \; , 
\end{align}
are invariant\footnote{I.e., their chiralspin is 0.} with respect to $SU(2)_{\textsc{cs}}$. In group-theoretical language, we have shown 
the multiplication rule $2 \otimes 2 = 3 \oplus 1$ for $SU(2)$. 

\begin{figure}
\centering
\includegraphics[angle=0,width=.95\linewidth]{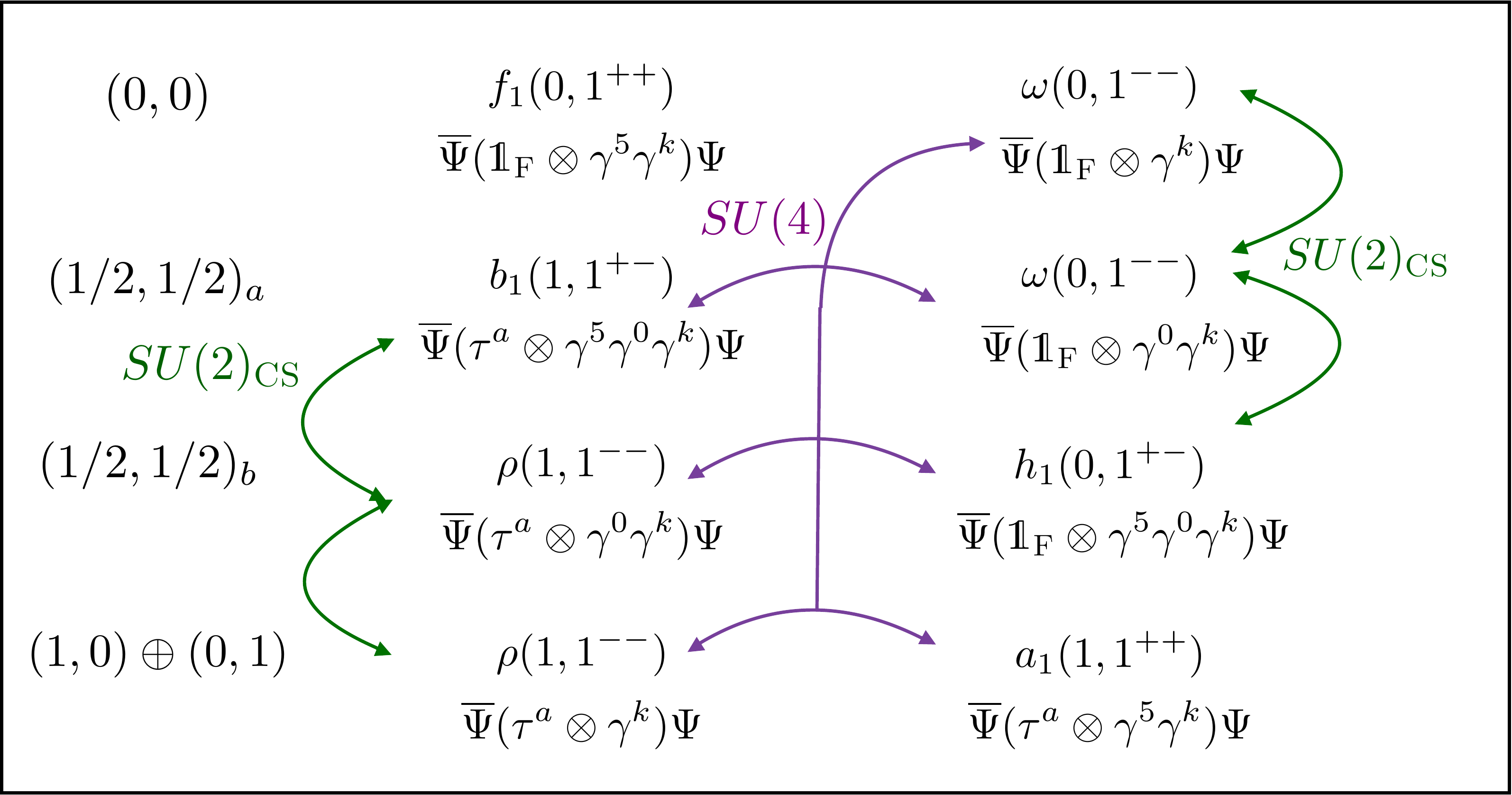}
\caption{{\sl} The $SU(2)_{CS}$ triplets are denoted by green lines; $f_1$ and $a_1$ are the  $SU(2)_{CS}$-singlets. The $SU(4)$ 15-plet is indicated by purple lines; $f_1$ is the $SU(4)$-singlet.}
 \label{Table2}
\end{figure}

The $SU(2)_{CS}$ triplets and singlets are shown in Fig.~ \ref{Table2}.

The $U(1)_A$ symmetry is contained in $SU(2)_{\textsc{cs}}$ as a subgroup.

Let us, at the end of this section, emphasize that 
the $SU(2)_{\textsc{cs}}$ symmetry is not a symmetry 
 of the QCD Lagrangian.
We apply
a $SU(2)_{\textsc{cs}}$ transformation on the fermion part of the QCD Lagrangian: 
\begin{align}
\label{Lagrangian}
 \overline{\Psi}' (\mathds{1}_{\textsc{F}} \otimes \gamma^{\mu} D_{\mu}) \Psi' = &\overline{\Psi} (\mathds{1}_{\textsc{F}} \otimes \gamma^0 D_0)  \Psi 
 -\nonumber \\
 -&\overline{\Psi}  (\mathds{1}_{\textsc{F}} \otimes \gamma^0) V^{\dagger} (\mathds{1}_{\textsc{F}} \otimes \gamma^0 \boldsymbol{\gamma} \cdot \boldsymbol{D}) 
 V \Psi \; . 
\end{align}
The $\gamma^0$-part is invariant under this transformation. 
The spacial part would only be invariant if $\overline{\chi}^{i} = \chi^{i} (i=1,2,3)$, see Eq.~(\ref{chi-def}), {\it i.e.}, both the left-
and right-handed fermions fulfilled 
the same Weyl equations, as intended by the symmetry.\footnote{For example,
this symmetry is manifestly violated by instantons. Only the left-handed
quark satisfies the Dirac equation with zero eigenvalue in the field of an instanton,
while only the right-handed quark produces a zero-mode in the field of
an anti-instanton, Refs.~\cite{i1,i2}.} 
An invariance can also be achieved by a spatial coupling in the Lagrangian of the form $\gamma^5 \gamma^k$, see Eq.~(\ref{invar1}).

\subsection{ $SU(4)$}
\label{SU4-chapter}
When we try to find a common algebra for the  $SU(2)_L \times SU(2)_R$ and the $SU(2)_{\textsc{cs}}$ symmetries,
we immediately arrive at the $\mathfrak{su}(4)$ algebra. This is due to the commutator
\begin{align}
 \left[ (\mathds{1}_{\textsc{F}} \otimes \Sigma^i), (\tau^a \otimes \Sigma^3)\right] = 2 \, i \epsilon^{i 3 k} (\tau^a \otimes \Sigma^k) \; ,
\end{align}
with $a=1,2,3$ and $i,k=1,2,3$. 
The $15$ matrices altogether 
\begin{align}
 \{(\tau^a \otimes \mathds{1}_D), (\mathds{1}_F \otimes \Sigma^i), (\tau^a \otimes \Sigma^i) \} \; ,
\end{align}
form the generators $T^l$ of the $\mathfrak{su}(4)$ algebra, satisfying the following commutation relations 
\begin{align}
\label{anti}
 [T^l, T^m] &= 2 i f^{l m n} T^n \; , f^{l m n} = \frac{1}{8 i} \mbox{Tr}[[T^l,T^m] T^n] \; , \\
\label{comm} 
 \{T^l, T^m \} &= 2 \delta^{l m} \mathds{1} + 2 d^{l m n} T^n \; , d^{l m n} = \frac{1}{8} \mbox{Tr}[\{T^l,T^m\} T^n] \; ,
\end{align}
with $f^{l m n}$ denoting the totally antisymmetric structure constants and $d^{l m n}$ a totally symmetric tensor, $l,m,n=1,2,...,15$.
The formula
\begin{align}
\label{relation}
 (\boldsymbol{\epsilon} \cdot \boldsymbol{T}) (\boldsymbol{\epsilon} \cdot \boldsymbol{T}) =
 \boldsymbol{\epsilon}^2 + \left( i f^{l m n}  + d^{l m n} \right) \epsilon^l \epsilon^m T^n \; ,
\end{align}
follows from the (anti)-commutation relations.

We denote this symmetry as
\begin{align}
\label{W-def}
\Psi \rightarrow  \Psi^{'} &= e^{i \boldsymbol{\epsilon} \cdot \boldsymbol{T}/2} \Psi \equiv W \, \Psi \; ,
\end{align}
with the fundamental vector $\Psi$ being 4-dimensional:
\begin{align}
\Psi =\begin{pmatrix} u_{\textsc{L}} \\ u_{\textsc{R}}  \\ d_{\textsc{L}}  \\ d_{\textsc{R}} \end{pmatrix} \; . 
\end{align}
The $SU(4)$ symmetry transformation  mixes both quark flavors and left-/right-handed components. For instance, a 
left-handed up quark now transforms as:
\begin{align}
 u_{\textsc{L}} \rightarrow a \cdot u_{\textsc{L}} + b \cdot u_{\textsc{R}} + c \cdot d_{\textsc{L}} + e \cdot d_{\textsc{R}} \; ,
\end{align}
with $a,b,c,d$ being functions of the rotation vector $\boldsymbol{\varepsilon}$. 
The new mixing, not present for $SU(2)_L \times SU(2)_R$ and $SU(2)_{\textsc{cs}}$, is between $u_{\textsc{L}}$ 
and $d_{\textsc{R}}$ (and accordingly for the other quark flavors). 

In principle the matrix $W$ could be written in linearized form (according to Eqs.~(\ref{U-closed}),(\ref{V-closed}))
\begin{align}
W= a_0 (\mathds{1}_{\textsc{F}} \otimes \mathds{1}_{\textsc{D}})  + a_1 (i \boldsymbol{\epsilon} \cdot \boldsymbol{T}/2) \; ,
\end{align}
with the coefficients $a_0, a_1$ being expressions of $f^{abc}$ and $d^{abc}$, see Ref.~\cite{W}. We perform an analytical
evaluation with \textit{Mathematica}, where we express $W$ by its spectral decomposition.
We calculate which fields (mesons) are connected via $SU(4)$ 
by transforming each interpolator in Fig.~\ref{Table1} with respect to $W$, Eq.~(\ref{W-def}). 
We arrive that the following interpolators get mixed via $W$:
\begin{align}
\label{W-mes}
 J_{(0,1^{--})}^{(0,0)} &\leftrightarrow J_{(0,1^{--})}^{(1/2,1/2)_a}  \leftrightarrow J_{(0,1^{+-})}^{(1/2,1/2)_b} 
 \leftrightarrow J_{(1,1^{--})}^{(1,0) \oplus (0,1)} \\ &\leftrightarrow J_{(1,1^{+-})}^{(1/2,1/2)_a} \leftrightarrow   J_{(1,1^{--})}^{(1/2,1/2)_b} 
 \leftrightarrow J_{(1,1^{++})}^{(1,0) \oplus (0,1)} \nonumber \; .
\end{align}
They form basis vectors for a dim$=15$ irreducible representation of $SU(4)$.
Hence, any of the currents above, when transformed with respect to $W$, Eq.~(\ref{W-def}), can be decomposed as
\begin{align}
&\overline{\Psi} (\Xi^{\alpha} \otimes \gamma^k) \Psi \cdot \mathcal{E}^{\alpha}_{(i)} + \overline{\Psi} (\Xi^{\alpha} \otimes \gamma^0 \gamma^k) \Psi \cdot \mathcal{F}^{\alpha}_{(i)} \\
 + &\overline{\Psi} ( \Xi^{\alpha} \otimes  \gamma^5 \gamma^0 \gamma^k) \Psi \cdot \mathcal{G}^{\alpha}_{(i)} +  \overline{\Psi} (\tau^a \otimes \gamma^5 \gamma^k) \Psi \cdot \mathcal{K}^a_{(i)} \; , 
\end{align}
where we used the compact notation $\Xi^{\alpha}=(\mathds{1}_{\textsc{F}}, \tau^a)$, ($\alpha=1,2,3,4$) 
and $\mathcal{E}^{\alpha}_{(i)}, \mathcal{F}^{\alpha}_{(i)}, \mathcal{G}^{\alpha}_{(i)}, \mathcal{K}^a_{(i)}$
are functions of the rotation parameter $\boldsymbol{\varepsilon}$ only. The index $i$ labels the interpolators.

In this decomposition the interpolator
\begin{align}
\label{invarSU4}
 J^{(0,0)}_{(0,1^{++})} &= \overline{\Psi} (\mathds{1}_{\textsc{F}} \otimes \gamma^5 \gamma^k) \Psi \; , 
\end{align}
is missing, because it is invariant with respect to $W$, Eq.~(\ref{W-def}),
{\it i.e.}, represents a singlet representation of $SU(4)$.
We have thus shown the following $SU(4)$ multiplication
rule: $\overline{4} \otimes 4 = 15 \oplus 1$. 
The $SU(4)$ singlet and 15-plet are shown in Fig.~ \ref{Table2}.

\subsection{Other transformations}
The $SU(2)_L \times SU(2)_R$  and $SU(2)_{\textsc{cs}}$ symmetries are 
two subgroups of $SU(4)$. The matrices
\begin{align}
 T^{a,i} = \{ (\tau^a \otimes \mathds{1}_D), (\tau^a \otimes \Sigma^i) \} \; , \quad i = 1,2 \; ,
\end{align}
with $\Sigma^1$ and $\Sigma^2$  given in Eq.~(\ref{sigma}),
generate two additional subgroups of $SU(4)$. 
The transformations
\begin{align}
\label{X}
\Psi \rightarrow  \Psi^{'} &= e^{i (\frac{\boldsymbol{\epsilon} \cdot \boldsymbol{\tau}}{2} \otimes \Sigma_1)} \Psi \equiv X \, \Psi \; , \\
\label{Y}
\Psi \rightarrow  \Psi^{'} &= e^{i (\frac{\boldsymbol{\epsilon} \cdot \boldsymbol{\tau}}{2} \otimes \Sigma_2)} \Psi \equiv Y \, \Psi \; . 
\end{align}
do not form closed subgroups but we denote them  for shortness as $SU(2)_X$ and $SU(2)_Y$. 
They can be expressed in closed form according to Eq.~(\ref{U-closed}) with $\gamma^0$ ($i \gamma^5 \gamma^0$) instead of $\gamma^5$ in Dirac space. 

The following left-right-handed quark flavors mix with $u_{\textsc{L}}$ via these symmetries:
\begin{align}
 u_{\textsc{L}} \rightarrow a \cdot u_{\textsc{L}} + b \cdot u_{\textsc{R}} + c \cdot d_{\textsc{R}} \, ,
\end{align}
which means that $u_{\textsc{L}}$ mixes with all $L/R$-quark flavors except $d_{\textsc{L}}$. The same is true 
for $u_{\textsc{R}}$, which mixes with all flavors except $d_{\textsc{R}}$. So the mixings 
of the chiral $SU(2)_L \times SU(2)_R $  symmetry, namely $u_{\textsc{L}} \leftrightarrow d_{\textsc{L}}, 
u_{\textsc{R}} \leftrightarrow d_{\textsc{R}}$, do not occur for these two 
$X$ and $Y$ transformations. 

We now identify which 
interpolators become connected.
We start with the transformation  $X$, Eq.~(\ref{X}), for which the following mixings occur:
\begin{align}
J_{(0,1^{--})}^{(1/2,1/2)_a} &\longleftrightarrow J_{(1,1^{--})}^{(1,0) \oplus (0,1)} \; ,  \\
J_{(1,1^{+-})}^{(1/2,1/2)_a} &\longleftrightarrow J_{(1,1^{++})}^{(1,0) \oplus (0,1)} \; , \\
J_{(1,1^{--})}^{(1/2,1/2)_b} &\longleftrightarrow J_{(0,1^{--})}^{(0,0)} \; .
\end{align}
The interpolators 
\begin{align}
 J_{(0,1^{++})}^{(0,0)} &= \overline{\Psi} (\mathds{1}_{\textsc{F}} \otimes \gamma^5 \gamma^k) \Psi \; , \\
 J_{(0,1^{+-})}^{(1/2,1/2)_b} &= \overline{\Psi} (\mathds{1}_{\textsc{F}} \otimes \gamma^5 \gamma^0 \gamma^k) \Psi \; ,
\end{align}
are invariant. 

Now we turn to the  transformation $Y$, Eq.~(\ref{Y}).
Here the particles with interpolators
\begin{align}
 J_{(0,1^{+-})}^{(1/2,1/2)_b} &\longleftrightarrow J_{(1,1^{--})}^{(1,0) \oplus (0,1)} \; ,  \\
 J_{(1,1^{--})}^{(1/2,1/2)_b} &\longleftrightarrow J_{(1,1^{++})}^{(1,0) \oplus (0,1)} \; , \\
J_{(1,1^{+-})}^{(1/2,1/2)_a} &\longleftrightarrow J_{(0,1^{--})}^{(0,0)} \; .
\end{align}
form doublets.
The interpolators 
\begin{align}
 J_{(0,1^{++})}^{(0,0)} &= (\mathds{1}_{\textsc{F}} \otimes \gamma^5 \gamma^k) \; , \\
J_{(0,1^{--})}^{(1/2,1/2)_a} &= (\mathds{1}_{\textsc{F}} \otimes \gamma^0 \gamma^k) \; , 
\end{align}
are invariant.

\begin{figure}
\centering
\includegraphics[angle=0,width=.95\linewidth]{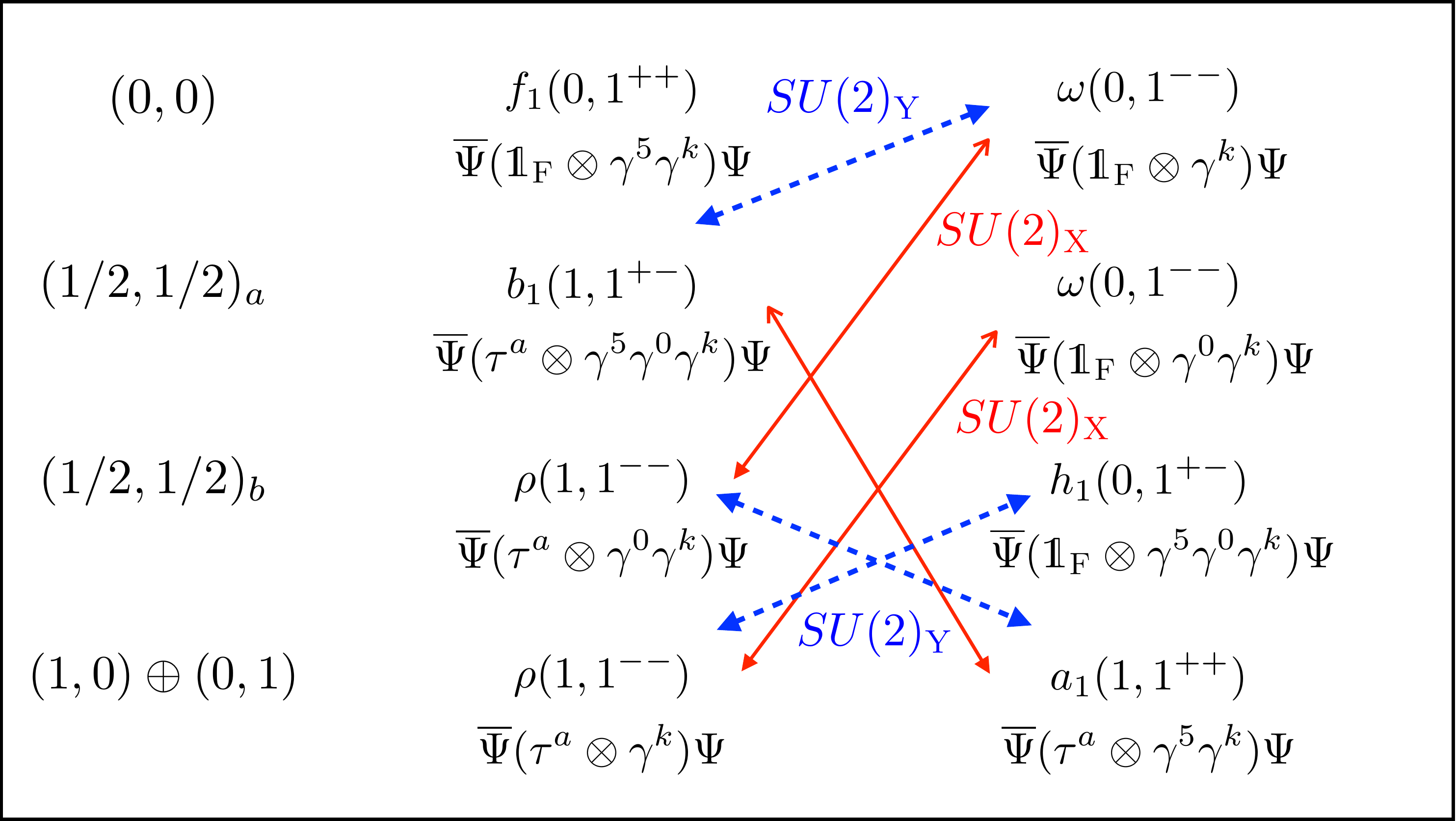}
\caption{\sl Symmetry  transformations $SU(2)_X$, Eq.~(\ref{X}) (red), $SU(2)_Y$, Eq.~(\ref{Y}) (dotted blue) 
connecting spin-1 meson fields. $SU(4)$}
 \label{Table3}
\end{figure}

To make our findings more transparent, in  Fig.~\ref{Table3} we show how the  transformations $SU(2)_X$ (red), $SU(2)_Y$ (dotted blue) 
connect the different mesons of spin-1. 

\subsection{Generalization to arbitrary spin}
The $SU(4)$ symmetry holds also for arbitrary spin $J \geq 1$,
because for any $J \geq 1$ one can construct interpolators with derivatives
that have exactly the same chiral transformation properties as those in Fig.~\ref{Table1},
see for details Ref.~\cite{G2}.

For even spins, $J=2n, n=1,2,...$ we have the following 15-plets
\begin{align}
 J_{(0,J^{++})}^{(0,0)} &\leftrightarrow J_{(0,J^{++})}^{(1/2,1/2)_a}  \leftrightarrow J_{(0,J^{-+})}^{(1/2,1/2)_b} 
 \leftrightarrow J_{(1,J^{++})}^{(1,0) \oplus (0,1)} \\ &\leftrightarrow J_{(1,J^{-+})}^{(1/2,1/2)_a} \leftrightarrow J_{(1,J^{++})}^{(1/2,1/2)_b} 
 \leftrightarrow J_{(1,J^{--})}^{(1,0) \oplus (0,1)} \nonumber \; ,
\end{align}
and for mesons with spin $J=2n-1$ we have
\begin{align}
 J_{(0,J^{--})}^{(0,0)} &\leftrightarrow J_{(0,J^{--})}^{(1/2,1/2)_a}  \leftrightarrow J_{(0,J^{+-})}^{(1/2,1/2)_b} 
 \leftrightarrow J_{(1,J^{--})}^{(1,0) \oplus (0,1)} \\ &\leftrightarrow J_{(1,J^{+-})}^{(1/2,1/2)_a} \leftrightarrow J_{(1,J^{--})}^{(1/2,1/2)_b}
  \leftrightarrow J_{(1,J^{++})}^{(1,0) \oplus (0,1)} \nonumber \; .
\end{align}
The $SU(4)$-singlets are  $J_{(0,J^{--})}^{(0,0)}$ for even spin and
$J_{(0,J^{++})}^{(0,0)}$ for odd spin.

For $J=0$ only the $(1/2,1/2)_a$ and $(1/2,1/2)_b$ chiral representations are possible and the symmetry group is $SU(2)_L \otimes SU(2)_R \times U(1)_A$.

\subsection{Generalization to three and $N_{\textsc{F}}$ flavors}
\label{SU3-chapter}
The three flavor-mesons are classified according to $SU(3)_L \otimes SU(3)_R$,
and fall into the irreducible representations $(1,1), (\overline{3},3) \oplus (3,\overline{3}), (8,1)\oplus (1,8)$.
The symmetry connecting the interpolators in these distinct irreducible representations is $SU(6)$ with the 
$35$ generators 
\begin{align}
 T^l = \{ (\lambda^a \otimes \mathds{1}_{\textsc{D}}), (\mathds{1}_{\textsc{F}} \otimes \Sigma^i), (\lambda^a \otimes \Sigma^i) \} \; ,
\end{align}
and $\lambda^a$ the Gell-Mann matrices ($a=1,\ldots,8$), $l=1,2,\ldots 35$.  
The fundamental vector $\Psi$ is six-dimensional and 
we have for fixed spin $J$ the multiplication rule: $\overline{6} \otimes 6 = 35 \oplus 1$. 
This can be further generalized to $N_{\textsc{F}}$ flavors, by simply replacing the Gell-Mann $\lambda^a$ matrices in $T^l$
with any other $\mathfrak{su}(N_{\textsc{F}})$-generators in flavor
space. We then arrive at the $(2 N_{\textsc{F}})^2-1$ generators of the $SU(2 N_{\textsc{F}})$ symmetry. 
All symmetry patterns derived in the above sections for two flavors, apply for three and $N_{\textsc{F}}$ flavors as well. 

\section{Implications}
We have mentioned in the introduction, that upon subtraction of the
lowest-lying Dirac modes from the valence quark propagators a degeneracy
of all  mesons from the 15-plet is observed. Clearly, this is not accidental
and reflects some inherent in QCD dynamics. 

A priori one expects that elimination of the quasi-zero modes should restore
the chiral $SU(2)_L \times SU(2)_R$ symmetry in hadrons since the quark condensate of
the vacuum is connected with the density of the quasi-zero modes via the
Banks-Casher relation \cite{BC}. However, not only  degeneracy patterns from
the groups $SU(2)_L \times SU(2)_R$ and $U(1)_A$ are seen, but a larger symmetry
$SU(4) \supset SU(2)_L \times SU(2)_R \times U(1)_A$. 

Naively one could assume that all the interesting nonperturbative physics
is removed with the low-lying modes and what remains  is some output from
perturbative interactions. Such a simple assumption can be ruled out, however.
In Ref. \cite{DGL1} we have proven that a system of free or weakly interacting
quarks in a box is not compatible with the degeneracy of the ground states with opposite spatial parity seen in our lattice results. Such a degeneracy necessarily implies that
we deal with the bound (confined) system of quarks. Secondly, within a perturbative approach only the $SU(2)_L \times SU(2)_R$ symmetry can be obtained, since it is a symmetry of the QCD Lagrangian and consequently of perturbation theory. Thirdly, the energy of the quark-antiquark system with perturbative interactions should be around two bare quark masses, and not of the order 1 GeV. 

The very fact that we observe a higher symmetry than the symmetry of the QCD Lagrangian does imply that we deal with a highly nontrivial nonperturbative
system. In Ref.~\cite{G1} it has been proven that there is no color-magnetic field
in this bound (confined) system. There is only a color-electric field that binds
the quarks. Here we present an alternative way to support this statement.
 
 From Eq.~(\ref{Lagrangian}) it follows that the colour-Coulomb part
 of the
 QCD Hamiltonian in Coulomb gauge ($J$ denotes the Faddeev-Popov determinant)
 \begin{equation}
 H_C = \frac{g^2}{2} \int J^{-1} \ \rho^a(\boldsymbol{x})  F^{ab}(\boldsymbol{x},\boldsymbol{y}) \, J \, \rho^b(\bf y) \; ,
 \label{inst}
\end{equation}
which describes the 
interaction between color charge densitites $\rho^a(\bf x)$ mediated by the color-Coulomb kernel
$F^{a b}(\bf x,\bf y)$ is invariant with respect to both $SU(2)_{CS}$ and $SU(4)$ transformations.
Therefore such a term survives in an $SU(4)$-symmetric hadron. It is the term which arises from the longitudinal part of the color-electric Yang-Mills Hamiltonian after resolving Gauss law. 

However, the coupling of quarks to spatial gluons
 \begin{equation}
H_T = -g \int d^3 x \, \Psi^\dag({\boldsymbol{x}}) \boldsymbol{\alpha} \cdot \boldsymbol{A}(\boldsymbol{x}) \, \Psi(\boldsymbol{x}) \; , 
 \end{equation}
is not $SU(2)_{CS}$- and $SU(4)$-symmetric and therefore its expectation value must vanish in the $SU(4)$-symmetric  hadron wave function. The only interaction left is the color-Coulomb part $H_C$, Eq.~(\ref{inst}). 
\footnote{It could be also seen
directly from (23): the $\gamma^0$ part of the interaction Lagrangian
represents the Coulombic 
interaction coming from Gauss law, while the spatial part is due to the interaction of the quark current
with spatial gluons  ${\bf j \cdot A}$, where colour-magnetic contributions can occur.}

We conclude that after reduction of the quasi-zero modes the only interaction left in the system
arises from the color-electric field components and there is no color-magnetic field contribution.
In the untruncated case the color-magnetic field contribution is still there,
but only through the quasi-zero modes. Indeed, e.g., the instanton fluctuations,
which do lead the quasi-zero modes \cite{S,DP},
do contain the magnetic field.

The instantaneous color-Coulomb
potential, given as the expectation value of  $F^{ab}$ in the gluon sector, is confining \cite{Z} and can be approximately obtained from the variational approach \cite{R}.

The  linear color-Coulomb potential implies that lines of the colour-electric field are squeezed into a flux-tube. A flux-tube between static quark
sources has been observed on the lattice, see Ref.~\cite{B} and references
therein. The observed $SU(4)$ symmetry of hadrons after the low-mode
elimination can be  consequently connected to the existence of the {\it dynamical} QCD string
and its energy \cite{G1}.

\section{Summary}
\label{Summary}
We have found a new $SU(4)$ symmetry of the bilinear quark-antiquark fields
of any spin $J \geq 1$ with non-exotic quantum numbers. This symmetry contains not only chiral transformations,
but also the left-right rotations of massless quarks. We have classified
interpolating fields according to different irreducible representations
of $SU(4)$ and its subgroups. These results are straightforwardly generalized
to $N_{\textsc{F}}$ massless flavors and the respective group is $SU(2N_{\textsc{F}})$.

The very fact that the correlation functions calculated with all operators 
from the 15-plet of $SU(4)$ in Ref. \cite{DGL2} upon subtraction of the quasi-zero modes of the Dirac operator
get indistinguishable, establishes the new $SU(4)$ symmetry of mesons after removal of the quasi-zero modes. This symmetry is higher than the symmetry of
the QCD Lagrangian and should be consequently considered as an emergent
symmetry. This symmetry implies the absence of magnetic interactions 
(of the color-magnetic field) in the system and might be interpreted as a manifestation of the
dynamical QCD string, Ref.~\cite{G1}.

An interesting question is whether the $f_1$ meson, that belongs to the
singlet representation of $SU(4)$, is degenerate or not with other $J=1$ mesons.
If yes, then there should exist a higher symmetry, that contains $SU(4)$
as a subgroup and that combines both the 15-plet and the singlet representation into a higher representation. 
It cannot be $U(4)$, because  a transition from $U(4)$ to its subgroup $SU(4)$ does not reduce the irreducible representations of $U(4)$ 
into a sum of irreducible representations of $SU(4)$. 

\begin{acknowledgements}
The authors acknowledge support by the Austrian Science Fund (FWF)
through the grant P26627-N27.  
\end{acknowledgements}

\begin{appendix}
\section{Basic Definitions and Conventions}
\label{App-def}
The chiral representation of $\gamma$-matrices enables us to write $\gamma^{\mu}$ in a compact notation:
\begin{align}
\label{chi-def}
 \gamma^{\mu} = \begin{pmatrix}
  0 & \chi^{\mu} \\
  \overline{\chi}^{\mu} & 0
 \end{pmatrix} \; , \quad \chi^{\mu} = \left(1, \boldsymbol{\chi}\right) \; , \quad \overline{\chi}^{\mu} = \left(1, - \boldsymbol{\chi}\right) \; . 
\end{align}
and the chirality matrix $\gamma^5$ is given as 
\begin{align}
 \gamma^5 = \begin{pmatrix}
  -1 & 0 \\
  0 & 1
 \end{pmatrix} \; ,
\end{align}
so that for a single flavor the quark field in L/R-components is given as
\begin{align}
\psi= \begin{pmatrix}
  \phi_L \\
  \phi_R   
 \end{pmatrix} \; .
\end{align}
Important for the construction of the meson symmetries are matrices of the form $M_{\textsc{F}} \otimes N_{\textsc{D}}$ with
$M$ and $N$ matrices in flavor and Dirac space, respectively. 
As an example, we construct:  
\begin{align}
  (\mathds{1}_{\textsc{F}} \otimes \gamma^k) &= \begin{pmatrix}
  0 & \chi^k & 0 & 0 \\
  -\chi^k & 0 & 0 & 0 \\
  0 & 0 & 0 & \chi^k \\
  0& 0 & -\chi^k & 0
 \end{pmatrix} \; , \\
 (\tau^1 \otimes \gamma^k) &= \begin{pmatrix}
  0 & 0 & 0 & \chi^k \\
  0 & 0 & -\chi^k & 0 \\
  0 & \chi^k & 0 & 0 \\
  -\chi^k & 0 & 0 & 0
 \end{pmatrix} \; ,  \\ 
  (\tau^2 \otimes \gamma^k) &= i \begin{pmatrix}
  0 & 0 & 0 &  -\chi^k \\
  0 & 0 & \chi^k & 0 \\
  0 & \chi^k & 0 & 0 \\
  - \chi^k & 0 & 0 & 0
 \end{pmatrix} \; ,  \\ 
   (\tau^3 \otimes \gamma^k) &= \begin{pmatrix}
  0 & \chi^k & 0 &  0 \\
  -\chi^k & 0 & 0 & 0 \\
  0 & 0 & 0 & -\chi^k \\
  0 & 0 & \chi^k & 0
 \end{pmatrix} \; .  
\end{align}

\end{appendix}

\end{document}